\documentclass[11pt]{article}
\usepackage{mathrsfs}
\usepackage[justification=centering]{caption}
\usepackage{booktabs}
\usepackage[numbers,sort&compress]{natbib}
\usepackage[colorlinks,
            linkcolor=blue,
            anchorcolor=green,
            citecolor=magenta
            ]{hyperref}
\usepackage{mathrsfs,amssymb,amsfonts}
\usepackage{graphicx}
\usepackage{amsmath,amssymb,latexsym,color}
\usepackage[mathscr]{eucal}
\usepackage[numbers]{natbib}
\newtheorem{thm}{Theorem}[section]
\newtheorem{definition}[thm]{Definition}

\newtheorem{lemma}[thm]{Lemma}

\newtheorem{example}{Example}
\newtheorem{remark}{Remark}[section]

\newcommand{\proof}{{\it Proof.\quad}}
\newcommand{\qed}{\hfill\Box\medskip}
\usepackage{CJK}
\usepackage{setspace}

\begin{document}

\title{\bf The $g$-Good-Neighbor Conditional Diagnosability of Locally Twisted Cubes}

\author
{Yulong Wei\quad Min Xu\footnote{\scriptsize Corresponding author. {\em E-mail address:} xum@bnu.edu.cn (M. Xu).}\\
{\footnotesize   \em  Sch. Math. Sci. {\rm \&} Lab. Math. Com. Sys., Beijing Normal University, Beijing, 100875, China}}
\date{}

\maketitle

\setlength{\baselineskip}{24pt}

\noindent {\bf Abstract}\quad In the work of Peng et al. in 2012, a new measure was proposed for fault diagnosis of systems: namely, $g$-good-neighbor conditional diagnosability, which requires that any fault-free vertex has at least $g$ fault-free neighbors in the system. In this paper, we establish the $g$-good-neighbor conditional diagnosability of locally twisted cubes under the PMC model and the MM$^*$ model. 

\noindent {\bf Keywords:}\quad PMC model; MM$^*$ model; Locally twisted cubes; Fault diagnosability
\vskip0.6cm

\section{Introduction}

With the size of multiprocessor systems increasing, processor failure is inevitable. Thus, to evaluate the reliability of multiprocessor systems, fault diagnosability has become an important metric. Many models have been proposed for determining a multiprocessor system's diagnosability. The PMC model was proposed by Preparata, Metze and Chien \cite{PMC} for fault diagnosis in multiprocessor systems. In the PMC model, all processors in the system under diagnosis can test one another. The MM model, proposed by Maeng and Malek \cite{MM}, assumes that a vertex in the system sends the same task to two of its neighbors and then compares their responses. Sengupta and Dahbura \cite{SD} further suggested a modification of the MM model, called the MM$^*$ model, in which each processor has to test two processors if the processor is adjacent to the latter two processors. Many researchers have applied the PMC model and the MM$^*$ model to identify faults in various topologies \big(see, for example, \cite{CCC,CH,CHS,HL,XTH,ZLX}\big).

The classical diagnosability for multiprocessor systems assumes that all the neighbors of any processor may fail simultaneously. However, the probability that this event occurs is very small in large-scale multiprocessor systems. In 2005, Lai et al. \cite{LTC} introduced conditional diagnosability under the assumption that all the neighbors of any processor in a multiprocessor system cannot be faulty at the same time. The conditional diagnosability of interconnection networks has been extensively investigated \big(see \cite{HCS,HL,XTH,Y,Z,ZQ}, etc.\big).

In 2012, Peng et al. proposed $g$-good-neighbor conditional diagnosability \cite{PLT}, which extended the concept of conditional diagnosability. This requires that every fault-free vertex has at least $g$ fault-free neighbors. Peng et al. \cite{PLT} studied the $g$-good-neighbor conditional diagnosability of the $n$-dimensional hypercube $Q_n$ under the PMC model. Since then, many researchers have studied this topic. For example, Wang et al. \cite{WLW,WGW} determined the $1,2$-good-neighbor diagnosability of the Cayley graph generated by transposition trees under the PMC model and the MM$^*$ model; Wang and Han \cite{WHM} determined the $g$-good-neighbor diagnosability of the $n$-dimensional hypercube $Q_n$ under the MM$^*$ model; Yuan et al. \cite{YLM,YLQ} established the $g$-good-neighbor diagnosability of the $k$-ary $n$-cubes under the PMC model and the MM$^*$ model; and Lin et al. \cite{LXW} determined the $g$-good-neighbor conditional diagnosability of arrangement graphs under the PMC model and the MM$^*$ model.

In this paper, we consider the $g$-good-neighbor conditional diagnosability of a well-known network, the $n$-dimensional locally twisted cube $LTQ_n$, under the PMC model and the MM$^*$ model. Our main results are listed below.

\bigskip

\noindent \textbf{Theorem \ref{main1}}~~
Let $n$ be an integer with $n\geq 4$. Then, the $g$-good-neighbor conditional diagnosability of $LTQ_n$ under the PMC model is
\begin{equation*}\label{}
t_g(LTQ_n)=
\left\{
  \begin{array}{lll}
   2^g(n-g+1)-1, & \hbox{$1\leq g\leq n-3$}; \\
    2^{n-1}-1, & \hbox{$n-2\leq g\leq n-1$.}
  \end{array}
\right.
\end{equation*}

\medskip

\noindent \textbf{Theorem \ref{main2}}~~
Let $n$ be an integer with $n\geq 5$. Then, the $g$-good-neighbor conditional diagnosability of $LTQ_n$ under the MM$^*$ model is
\begin{equation*}\label{}
t_g(LTQ_n)=
\left\{
  \begin{array}{lll}
   2^g(n-g+1)-1, & \hbox{$1\leq g\leq n-3$}; \\
    2^{n-1}-1, & \hbox{$n-2\leq g\leq n-1$.}
  \end{array}
\right.
\end{equation*}

The rest of this paper is organized as follows. Section \ref{3} introduces some terminology and preliminaries. Our main results are given in Section \ref{4}. Finally, Section \ref{6} concludes the paper.

\section{Terminology and preliminaries}\label{3}

An undirected simple graph $G =\big(V(G), E(G)\big)$ is used to represent a system (or a network) where each vertex represents a processor and each edge represents a link. A {\em subgraph} $H$ of $G$ is a graph with $V(H)\subseteq V(G)$, $E(H) \subseteq E(G)$ and the endpoints of every edge in $E(H)$ belonging to $V(H)$. For an arbitrary subset $F\subseteq V(G)$, we use $G-F$ to denote the graph obtained by removing all the vertexes in $F$ from $G$. Given a nonempty vertex subset $V'$ of $V(G)$, the {\em induced subgraph} by $V'$ in $G$, denoted by $G[V']$, is a graph in which the vertex set is $V'$ and the edge set is the set of all the edges of $G$ with both endpoints in $V'$. For a given vertex $v$, we define the {\em neighborhood} $N_G(v)$ of $v$ in $G$ to be the set of vertices adjacent to $v$. The {\em degree} of vertex $v$, denoted by $d_G(v)$, is the number of vertices in $N_G(v)$. The {\em minimum degree} of a graph $G$, denoted by $\delta(G)$, is $\min\limits_{v\in V(G)} d_G(v)$. A graph $G$ is $k$-{\em regular} if $d_G(v)=k$ for any $v\in V$. For a given set $A\subseteq G$, we denote by $N_G(A)$ the set $\big(\bigcup _{v\in V(A)}N_G(v)\big)-V(A)$. For neighborhoods and degrees, we omit the subscripts of the graphs when no confusion arises. The {\em symmetric difference} of two sets $F_1$ and $F_2$ is defined as the set $F_1\bigtriangleup F_2$ $=(F_1-F_2)\cup (F_2-F_1)$. Please refer to \cite{BMG} for graph-theoretical terminology and notation undefined here.

Now, we focus on the $n$-dimensional locally twisted cube $LTQ_n$.

The $n$-bit binary string is denoted by $\{0, 1\}^n$. Let $``\bigoplus"$ represent modulo $2$ addition. For any two binary bits $u$, $v\in\{0, 1\}$, let $u\bigoplus v$ be the sum modulo $2$ of $u$ and $v$, and $\overline{u}=u\bigoplus 1$. The formal definition of $LTQ_n$ is provided as follows.

\begin{definition}{\rm\cite{YDM5}}\label{D1}
Let $n$ be a positive integer. The locally twisted cube $LTQ_n$ of dimension $n$ has $2^n$ vertices, each labeled by
an $n$-bit binary string $u_{n-1}\ldots u_1u_0$. $LTQ_n$ is defined recursively as follows:

{\rm(1)} $LTQ_2$ is a graph comprising four nodes, labeled $00, 01, 10$ and $11$, which are connected by four edges
$(00, 01), (01, 11), (11, 10)$ and $(10, 00)$. 

{\rm(2)} For $n \geq 3$, $LTQ_n$ is built from two disjoint copies of $LTQ_{n-1}$ according to the following steps: Let $LTQ_{n-1}^0$ denote the graph obtained from one copy of $LTQ_{n-1}$ by prefixing the label of each node with $0$. Let $LTQ_{n-1}^1$ denote the graph obtained from the other copy of $LTQ_{n-1}$ by prefixing the label of each node with $1$. Each node $0x_{n-2}x_{n-3}\ldots x_0$ of $LTQ_{n-1}^0$ is connected to the node $1(x_{n-2}\bigoplus x_0)x_{n-3}\ldots x_0$ of $LTQ_{n-1}^1$ by an edge.
\end{definition}

\begin{figure}[hptb]
  \centering
  \includegraphics[width=10cm]{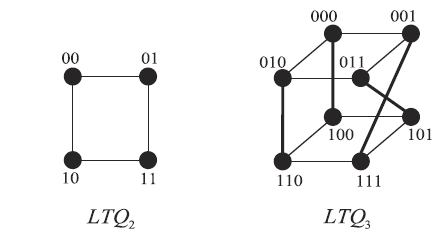}\\
  \caption{ $LTQ_2$ and $LTQ_3$
}\label{f2}
\end{figure}

According to Definition \ref{D1}, Figure \ref{f2} illustrates $LTQ_2$ and $LTQ_3$. An alternative definition of $LTQ_n$ is provided in the following non-recursive fashion:

\begin{definition}{\rm\cite{YDM}}\label{D2}
Let $n$ be a positive integer. The locally twisted cube $LTQ_n$ of dimension $n$ has $2^n$ vertices, each labeled by
an $n$-bit binary string $u_{n-1}\ldots u_1u_0$. Any two nodes $u=u_{n-1}\ldots u_1u_0$ and $v = v_{n-1}\ldots v_1v_0$ of $LTQ_n$ are adjacent if and only if one of the following conditions is satisfied.

{\rm(1)} There is an integer $k$, $2 \leq k \leq n-1$, such that

~~~~{\rm(a)} $u_k = \overline{v}_k$,

~~~~{\rm(b)} $u_{k-1} = v_{k-1}\bigoplus u_0$, and

~~~~{\rm(c)} all the remaining bits of $u$ and $v$ are identical.

{\rm(2)} $u_k = \overline{v}_k$ for some $k\in \{0, 1\}$ and $u_r = v_r$, where $r = 2, 3, \ldots , n-1$.
\end{definition}


Now we introduce two models for fault diagnosis.

In the PMC model, all processors in the system under diagnosis can test one another. The set of tests can be represented by a directed graph $G = (V, E)$, in which each vertex represents a processor, and an edge $(u, v)$ indicates that the processor $u$ has tested processor $v$. The outcome of processor $u$ testing processor $v$ is denoted by $\sigma(u, v)$, where
\begin{equation*}\label{}
\sigma(u, v)=
\left\{
  \begin{array}{lll}
   0, & \hbox{$\{u, v\}\cap F=\emptyset$}; \\
   1, & \hbox{$u\notin F$, $v\in F$}; \\
    0~{\rm or}~1, & \hbox{$u\in F$,}
  \end{array}
\right.
\end{equation*}
where $F$ is the set of faulty processors.

In the MM$^*$ model, a processor executes comparisons for any pair of its neighboring processors. A graph $G =(V, E)$ is used to represent a system, where each vertex represents a processor and each edge represents a link. Assign a task to each vertex. The vertex $w$ is a comparator of a pair of processors $\{u, v\}$ if $(u, w)\in E$ and $(v, w)\in E$. The outcome of this comparison is denoted by $\sigma\big((u, v)_w\big)$, where
\begin{equation*}\label{}
\sigma\big((u, v)_w\big)=
\left\{
  \begin{array}{lll}
   0, & \hbox{$\{u, v, w\}\cap F=\emptyset$}; \\
   1, & \hbox{$w\notin F$, $\{u, v\}\cap F\neq\emptyset$}; \\
    0~{\rm or}~1, & \hbox{$w\in F$,}
  \end{array}
\right.
\end{equation*}
where $F$ is the set of faulty processors.

The collection of all outcomes is called a {\em syndrome} $\sigma$. The diagnosis problem involves using the syndrome to determine the status (faulty or fault free) of each processor in the system. For a given syndrome $\sigma$, a subset $F\subseteq V$ is said to {\em be consistent with} $\sigma$ if the syndrome $\sigma$ can be produced from the faulty set $F$. In concrete terms, in the PMC model, $F$ is said to be consistent with $\sigma$ if the syndrome $\sigma$ can be produced from the situation that, for any $(u, v) \in E$ such that $u \notin F$, $\sigma(u, v)=1$ if and only if $v \in F$. In the MM$^*$ model, $F$ is said to be consistent with $\sigma$ if the syndrome $\sigma$ can be produced from the situation that, for any $(u, w)\in E$ and $(v, w)\in E$ such that $w \notin F$, $\sigma\big((u, v)_w\big)=1$ if and only if $\{u, v\}\cap F\neq\emptyset$. Therefore, on the one hand, a faulty set $F$ may produce a number of different syndromes. On the other hand, different faulty sets may produce the same syndrome. Define $\sigma (F)=\{\sigma \mid F ~{\rm is~ consistent~ with}~ \sigma\}$. Two distinct sets $F_1$, $F_2 \subseteq V$ are said to be indistinguishable if $\sigma(F_1)\cap \sigma(F_2)\neq\emptyset$; otherwise, $F_1$ and $F_2$ are said to be distinguishable. We say that $(F_1, F_2)$ is an indistinguishable pair if $\sigma(F_1)\cap \sigma(F_2)\neq\emptyset$; else, $(F_1, F_2)$ is a distinguishable pair.

The following lemmas give necessary and sufficient conditions for a pair of sets to be distinguishable under the PMC model and the MM$^*$ model. 

\begin{figure}[hptb]
 \centering
  \includegraphics[width=8cm]{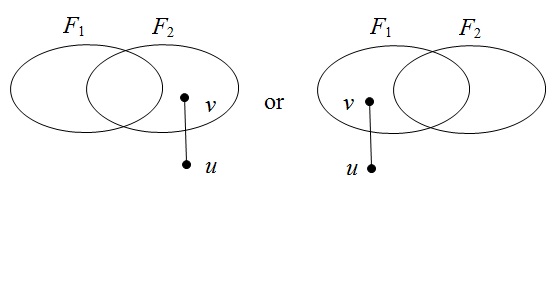}\\
  \caption{ The illustration of a distinguishable pair $(F_1, F_2)$ under the PMC model.
}\label{L11}
\end{figure}

\begin{figure}[hptb]
  \centering
  \includegraphics[width=8cm]{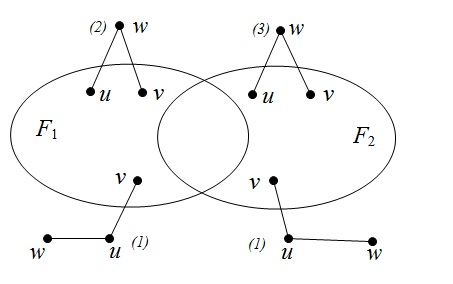}\\
  \caption{ The illustration of distinguishable sets $F_1$ and $F_2$ under the MM$^*$ model.
}\label{L22}
\end{figure}

\begin{lemma}{\rm\cite{DM}}\label{L01}
For any two distinct sets $F_1$, $F_2 \subseteq V$, $(F_1, F_2)$ is a distinguishable pair if and only if there exists a vertex $u\in V-(F_1\cup F_2)$ and there exists a vertex
$v\in F_1\bigtriangleup F_2$ such that $(u, v)\in E$ {\rm(}See Figure \ref{L11} {\rm)}.
\end{lemma}
\begin{lemma}{\rm\cite{SD}}\label{L02}
Let $G =(V, E)$ be a graph. For any two distinct sets $F_1$, $F_2 \subseteq V$, $F_1$ and $F_2$ are distinguishable under the MM$^*$ model if and only if any one of the following conditions is satisfied {\rm(}See Figure \ref{L22} {\rm)}:

{\rm (1)} There are two vertices $u, w\in V-(F_1\cup F_2)$ and there is a vertex $v\in F_1\bigtriangleup F_2$
such that $(u, v) \in E$ and $(u, w)\in E$.

{\rm (2)} There are two vertices $u, v\in F_1-F_2$ and there is a vertex $w\in V-(F_1\cup F_2)$
such that $(u, w) \in E$ and $(v, w)\in E$.

{\rm (3)} There are two vertices $u, v\in F_2-F_1$ and there is a vertex $w\in V-(F_1\cup F_2)$
such that $(u, w) \in E$ and $(v, w)\in E$.
\end{lemma}

Next, we introduce the diagnosability, conditional diagnosability, $R_g$-connectivity and $g$-good-neighbor conditional diagnosability of a system in the following statements.
\begin{definition}{\rm\cite{DM}}\label{D3}
A system of $n$ processors is $t$-diagnosable if all faulty processors can be detected without replacement, provided that the number of faults does not exceed $t$. The diagnosability $t(G)$ of system $G=(V, E)$ is the maximum value of $t$ such that $G$ is $t$-diagnosable.
\end{definition}

The diagnosability of multiprocessor systems, as defined above, assumes that all neighbors of any processor may fail simultaneously. However, the probability that all the neighbors of a processor fail is very small. In 2005, Lai et al. \cite{LTC} introduced conditional diagnosability under the assumption that all the neighbors of any processor in a multiprocessor system cannot be faulty at the same time.

\begin{definition}{\rm\cite{LTC}}\label{D4} System $G =(V, E)$ is conditionally $t$-diagnosable if $G$ is $t$-diagnosable, provided that for any processor $v \in V$, the set of faults does not contain the neighborhood $N(v)$ as a subset. The conditional diagnosability $t_c(G)$ of graph $G$ is the maximum value of $t$ such that $G$ is conditionally $t$-diagnosable.
\end{definition}

Inspired by the concept of conditional diagnosability, Peng et al. \cite{PLT} proposed $g$-good-neighbor conditional diagnosability in 2012, which extended the concept of conditional diagnosability.

\begin{definition}{\rm\cite{PLT,YLM}}\label{D5} A faulty set $F\subseteq V$ is called a $g$-good-neighbor conditional faulty set if $|N(v) \cap (V-F)|\geq g$ for each node $v$ in $V-F$. A $g$-good-neighbor conditional cut of a graph $G$
is a $g$-good-neighbor conditional faulty set $F$ such that $G-F$ is disconnected. The minimum
cardinality of $g$-good-neighbor cuts is said to be the $R_g$-connectivity of $G$, denoted
by $\kappa^g(G)$.
\end{definition}

\begin{definition}{\rm\cite{PLT}}\label{D6}
A system $G =(V, E)$ is $g$-good-neighbor conditional $t$-diagnosable if $G$ is $t$-diagnosable, provided that every faulty set is a $g$-good-neighbor conditional faulty set. The $g$-good-neighbor conditional diagnosability $t_g(G)$ of $G$ is the maximum value of $t$ such that $G$ is $g$-good-neighbor conditionally $t$-diagnosable.
\end{definition}

Thus, the following lemmas give necessary and sufficient conditions for a system to be $g$-good-neighbor $t$-diagnosable under the PMC model and under the MM$^*$ model. 
\begin{lemma}{\rm\cite{PLT,SD}}\label{L1}
A system $G =(V, E)$ is $g$-good-neighbor $t$-diagnosable under the
PMC model if and only if there is an edge $(u, v)\in E$ with $u\in V-(F_1\cup F_2)$ and
$v\in F_1\bigtriangleup F_2$ for each distinct pair of $g$-good-neighbor conditional faulty sets $F_1$ and $F_2$ of $V$ with $|F_1|\leq t$ and $|F_2|\leq t$ {\rm(}See Figure \ref{L11} {\rm)}.
\end{lemma}

\begin{lemma}{\rm\cite{DM,YLM}}\label{L2}
A system $G =(V, E)$ is $g$-good-neighbor $t$-diagnosable under the
MM$^*$ model if and only if each distinct pair of $g$-good-neighbor conditional faulty sets $F_1$ and
$F_2$ of $V$ with $|F_1|\leq t$ and $|F_2|\leq t$ satisfies one of the following conditions {\rm(}See Figure \ref{L22} {\rm)}:

{\rm (1)} There are two vertices $u, w\in V-(F_1\cup F_2)$ and there is a vertex $v\in F_1\bigtriangleup F_2$
such that $(u, v) \in E$ and $(u, w)\in E$.

{\rm (2)} There are two vertices $u, v\in F_1-F_2$ and there is a vertex $w\in V-(F_1\cup F_2)$
such that $(u, w) \in E$ and $(v, w)\in E$.

{\rm (3)} There are two vertices $u, v\in F_2-F_1$ and there is a vertex $w\in V-(F_1\cup F_2)$
such that $(u, w) \in E$ and $(v, w)\in E$.
\end{lemma}

Work related to the $R_g$-connectivity for special networks and small values of $g$ can be found in the literature \big(see, for example, \cite{CLY,HHL,LX,WH,WZ,YM}\big). The following lemmas are very useful for proving our main results. 

\begin{lemma}{\rm\cite{PLT}}\label{L3}
For any given graph $G$, if $g\leq g'$, then $t_g(G)\leq t_{g'}(G)$ under the PMC model and MM$^*$ model.
\end{lemma}

\begin{lemma}{\rm\cite{WGW}}\label{L8}
For any given graph $G$, $t(G)=t_0(G)$ under the PMC model and MM$^*$ model.
\end{lemma}

\begin{lemma}{\rm\cite{FZ}}\label{L7}
For any positive integer $n$, there is no cycle of length $3$ in the locally twisted cube $LTQ_n$.
\end{lemma}


\begin{lemma}{\rm\cite{WH}}\label{L5}
Let $H$ be a subgraph of $LTQ_n$. If $\delta(H)=g$, then $|V(H)|\geq 2^g$, where $0 \leq g \leq n$ and $n\geq 2$.
\end{lemma}

\begin{lemma}{\rm\cite{WH}}\label{L6}
For an $n$-dimensional locally twisted cube $LTQ_n$, $\kappa^g(LTQ_n)=2^g(n-g)$ if $n\geq g + 2$.
\end{lemma}

\section{Main Results}\label{4}
In this section, we will give the proofs of our main results.

Let $u_i=0~{\rm or}~1$ and $\alpha_i$ be a positive integer for $i=1,\ldots,s$. Let $v_i=0~{\rm or}~1$ and $\beta_i$ be a positive integer for $i=1,\ldots,t$. We use $u_1^{\alpha_1}\ldots u_s^{\alpha_s}X^gv_1^{\beta_1}\ldots v_t^{\beta_t}$ to denote the vertex set $$\{\underbrace{u_1\ldots u_1}_{\alpha_1}\ldots \underbrace{u_s\ldots u_s}_{\alpha_s}x_g\ldots x_1\underbrace{v_1\ldots v_1}_{\beta_1}\ldots \underbrace{v_t\ldots v_t}_{\beta_t}\mid x_i=0~{\rm or}~1~{\rm for}~1\leq i\leq g\}. $$
Specifically, we denote the vertex set $$\{\underbrace{0\ldots 0}_{n-g-1}x_{g}\ldots x_10\mid x_i=0~{\rm or}~1~{\rm for}~1\leq i\leq g\}$$ by $0^{n-g-1}X^g0$. We can also denote $\underbrace{0 \ldots 0}_n$ and $\underbrace{1 \ldots 1}_n$ by $0^n$ and $1^n$, respectively.

By Lemma \ref{L8}, $t(LTQ_n)=t_0(LTQ_n)$, where $t(LTQ_n)$ is the classical diagnosability of $LTQ_n$. We will consider $g\geq1$ in the following.

First, we consider the upper bound of the $g$-good-neighbor conditional diagnosability of the $n$-dimensional locally twisted cube $LTQ_n$.

\begin{figure}[hptb]
  \centering
  \includegraphics[width=8cm]{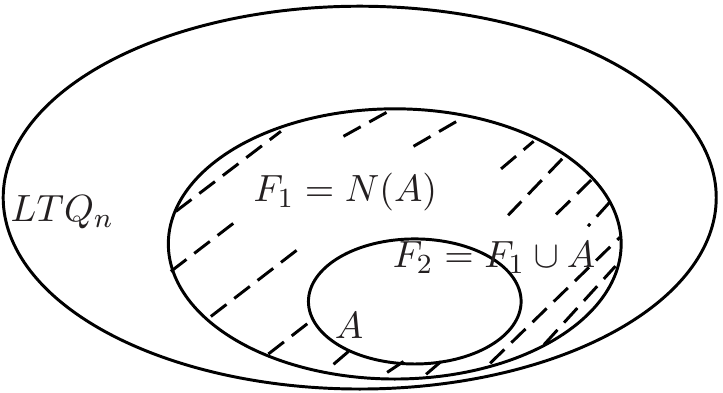}\\
  \caption{ The illustration of $F_1$ and $F_2$.
}\label{P11}
\end{figure}

\begin{lemma}\label{p1}
Let $n$ and $g$ be integers with $n\geq4$ and $1\leq g\leq n-3$. Then, we have $t_g(LTQ_n)\leq 2^g(n-g+1)-1$ under the PMC model and the MM$^*$ model.
\end{lemma}
\proof
For $1 \leq g\leq n-3$, let $A=0^{n-g-1}X^g0$, $F_1=N(A)$, and $F_2=F_1\cup A$. By Definition \ref{D2},
we note that $$F_1=10^{n-g-2}X^g0\bigcup 010^{n-g-3}X^g0\bigcup \ldots\bigcup 0^{n-g-2}1X^g0\bigcup 0^{n-g-1}X^g1. $$ Then, we have $|A|=2^g$, $|F_1|=2^g(n-g)$ and $|F_2|=2^g(n-g+1)$. Note that $A=F_1\bigtriangleup F_2$ and $F_1=N(A)$ (see Figure \ref{P11}). Since there is no edge between $F_1\bigtriangleup F_2$ and $V(LTQ_n)-(F_1\cup F_2)$, by Lemma \ref{L01} and Lemma \ref{L02}, we conclude that $F_1$ and $F_2$ are indistinguishable under the PMC model and the MM$^*$ model.

Now, we verify that both $F_1$ and $F_2$ are $g$-good-neighbor conditional faulty sets. 

Suppose $u=u_{n-1}u_{n-2}\ldots u_1u_0\in V(LTQ_n)-F_2$. If $u \notin N(F_2)$, then $N(u)\cap F_2=\emptyset$. If $u \in N(F_2)$, then $u \in N(F_1)$ owing to the fact that $F_2=F_1\cup A$ and $F_1=N(A)$. Thus, $u$ has three forms, i.e., $$0^p10^q10^{n-g-p-q-3}X^g0, $$ $$0^p110^{n-g-p-3}X^g1, $$ {\rm and}$$ 0^p10^{n-g-p-2}X^g1. $$ If $u \in 0^p10^q10^{n-g-p-q-3}X^g0$, then $$N(u)\cap F_2=\{0^{p+q+1}10^{n-g-p-q-3}u_g\ldots u_10, 0^p10^{n-g-p-2}u_g\ldots u_10\}. $$ If $u \in 0^p110^{n-g-p-3}X^g1$, then $$N(u)\cap F_2=\{0^{n-g-1}u_g\ldots u_11\}. $$ If $u \in 0^p10^{n-g-p-2}X^g1$, then
\begin{equation*}\label{}
N(u)\cap F_2=
\left\{
  \begin{array}{lll}
   \{0^p10^{n-g-p-2}u_g\ldots u_10\}, & \hbox{if $0\leq p\leq n-g-3$}; \\
    \{0^{n-g-2}1u_g\ldots u_10, 0^{n-g-1}\overline{u}_g\ldots u_11\}, & \hbox{if $p=n-g-2$.}
  \end{array}
\right.
\end{equation*}

\noindent Since $LTQ_n$ is $n$-regular, $|N(u)-F_2|\geq n-2\geq g$. Thus, $F_2$ is a $g$-good-neighbor conditional faulty set.

Suppose $u=u_{n-1}u_{n-2}\ldots u_1u_0\in V(LTQ_n)-F_1$. If $u \in V(LTQ_n)-F_2$, then we obtain the desired result by the same proof as above. If $u \notin V(LTQ_n)-F_2$, then $u \in A=0^{n-g-1}X^g0$. Note that $$N(u)\cap A=\{0^{n-g-1}\overline{u}_g\ldots u_10, 0^{n-g-1}u_g\overline{u}_{g-1}\ldots u_10,\ldots, 0^{n-g-1}u_g\ldots \overline{u}_10\}. $$Thus, $|N(u)\cap A|=g$, so $F_1$ is a $g$-good-neighbor conditional faulty set.

By the above discussion, we obtain the result. $\qed$

\begin{lemma}\label{p2}
Let $n$ and $g$ be integers with $n\geq4$ and $1\leq g\leq n-3$. Then, we have $t_g(LTQ_n)\leq 2^{n-1}-1$ under the PMC model and the MM$^*$ model.
\end{lemma}
\proof
Let $F_1=V(LTQ_{n-1}^0)$ and $F_2=V(LTQ_{n-1}^1)$. Then, $|F_1|=|F_2|=2^{n-1}$. Since $LTQ_{n-1}^0=LTQ_n-F_2$ and $LTQ_{n-1}^1=LTQ_n-F_1$, both $F_1$ and $F_2$ are $(n-1)$-good-neighbor conditional faulty sets. Note that $n-2\leq g\leq n-1$. Thus, by Definition \ref{D5}, both $F_1$ and $F_2$ are $g$-good-neighbor conditional faulty sets. Since $F_1\cup F_2=V(LTQ_n)$, i.e., $V(LTQ_n)-(F_1\cup F_2)=\emptyset$, by Lemma \ref{L01} and Lemma \ref{L02}, $F_1$ and $F_2$ are indistinguishable under the PMC model and the MM$^*$ model.

Thus, we obtain the result. $\qed$

Next, we consider the lower bound of the $g$-good-neighbor conditional diagnosability of the $n$-dimensional locally twisted cube $LTQ_n$ under the PMC model and the MM$^*$ model separately.
\begin{lemma}\label{p3}
Let $n$ and $g$ be integers with $n\geq4$ and $1\leq g\leq n-3$. Then, we have $t_g(LTQ_n)\geq 2^g(n-g+1)-1$ under the PMC model.
\end{lemma}
\proof Suppose that $F_1$ and $F_2$ are any two distinct $g$-good-neighbor conditional faulty sets and they are indistinguishable. We will prove the lemma by showing that $|F_1|\geq 2^g(n-g+1)$ or $|F_2|\geq 2^g(n-g+1)$.

If $V(LTQ_n)=F_1\cup F_2$, then $2^n = |V(LTQ_n)|$ $=|F_1\cup F_2|=|F_1| + |F_2|-|F_1\cap F_2|\leq |F_1|+|F_2|$, so $|F_1|\geq 2^{n-1}\geq 2^g(n-g+1)$ or $|F_2|\geq 2^{n-1}\geq 2^g(n-g+1)$ for $1\leq g\leq n-3$.

Now, we suppose $V(LTQ_n)\neq F_1\cup F_2$. Since $F_1$ and $F_2$ are indistinguishable, there are no edges between $V(LTQ_n)-(F_1\cup F_2)$ and $F_1\bigtriangleup F_2$ by Lemma \ref{L01}. Note that $F_1$ and $F_2$ are both $g$-good-neighbor conditional faulty sets. We know that $F_1\cap F_2$ is a $g$-good-neighbor conditional cut. By Lemma \ref{L6}, we obtain that $|F_1\cap F_2|\geq 2^g(n-g)$. Since $F_1\neq F_2$, without loss of generality, we assume that $F_2-F_1\neq \emptyset$. Since $F_1$ is a $g$-good-neighbor conditional faulty set, any vertex in $F_2-F_1$ has at least $g$ neighbors in $F_2-F_1$. By Lemma \ref{L5}, we have $|F_2-F_1|\geq 2^g$. Hence, $|F_2|=|F_2-F_1|+|F_1\cap F_2|\geq 2^g(n-g)+2^g=2^g(n-g+1)$.

This completes the proof of Lemma \ref{p3}. $\qed$

\begin{lemma}\label{p4}
Let $n$ and $g$ be integers with $n\geq5$ and $1\leq g\leq n-3$. Then, we have $t_g(LTQ_n)\geq 2^g(n-g+1)-1$ under the MM$^*$ model.
\end{lemma}
\proof
Suppose that $F_1$ and $F_2$ are any two distinct $g$-good-neighbor conditional faulty sets and they are indistinguishable. We will prove the lemma by showing that $|F_1|\geq 2^g(n-g+1)$ or $|F_2|\geq 2^g(n-g+1)$.

If $V(LTQ_n)=F_1\cup F_2$, then $2^n = |V(LTQ_n)|$ $=|F_1\cup F_2|=|F_1| + |F_2|-|F_1\cap F_2|\leq |F_1|+|F_2|$, so $|F_1|\geq 2^{n-1}\geq 2^g(n-g+1)$ or $|F_2|\geq 2^{n-1}\geq 2^g(n-g+1)$ for $1\leq g\leq n-3$. 

Now, we suppose $V(LTQ_n)\neq F_1\cup F_2$. Since $F_1\neq F_2$, without loss of generality, we assume that $F_2-F_1\neq \emptyset$. To prove this lemma, we consider two cases as follows.

{\em Case 1.} $2 \leq g\leq n-3$.

We shall show that there is no edge between $F_1\bigtriangleup F_2$ and $V(LTQ_n)-(F_1\cup F_2)$. Otherwise, there exists an edge $uv\in E(LTQ_n)$, where $u\in F_2-F_1$ and $v\in V(LTQ_n)-(F_1\cup F_2)$. Since $F_1$ is a $g$-good-neighbor conditional faulty set with $g\geq2$, $v$ has at least two neighbors in $LTQ_n-F_1$. Thus, $v$ has a neighbor $w~(w\neq u)$ in $F_2-F_1$ or $V(LTQ_n)-(F_1\cup F_2)$, which contradicts Lemma \ref{L02}. Note that $F_1$ and $F_2$ are $g$-good-neighbor conditional faulty sets. Thus, $F_1\cap F_2$ is a $g$-good-neighbor conditional cut of $LTQ_n$, so $|F_1\cap F_2|\geq 2^g(n-g)$ by Lemma \ref{L6}. Since $\delta (LTQ_n[F_2-F_1])\geq g$, we have $|F_2-F_1|\geq 2^g$ by Lemma \ref{L5}.  Therefore, $|F_2|=|F_1\cap F_2|+|F_2-F_1|\geq 2^g(n-g)+ 2^g=2^g(n-g+1)$.

{\em Case 2.} $g=1$.

In this case, we have $2^g(n-g+1)=2n$.

If $|F_2\cap F_1|\geq 2n-1$, then $|F_2|=|F_2-F_1|+|F_1\cap F_2|\geq 1 + (2n-1)=2n$.

Now, we suppose $|F_2\cap F_1|\leq 2n-2$. Let $W_1, \ldots, W_k$ be the components of $LTQ_n-(F_1\cup F_2)$ such that $|V(W_1)|\leq \ldots\leq |V(W_k)|$, where $k\geq 1$. For any component $W_i$, if $|V(W_i)|\geq 2$, then there is no edge between $W_i$ and $F_1\bigtriangleup F_2$. Otherwise, it contradicts the fact that $F_1$ and $F_2$ are indistinguishable. Let $W=\bigcup\limits_{\stackrel{|V(W_i)|=1}{1\leq i\leq k}}V(W_i)$.

If $W=\emptyset$, then $|V(W_i)|\geq 2$ for $1\leq i\leq k$. Note that $F_1$ and $F_2$ are $1$-good-neighbor conditional faulty sets. Thus, $F_1\cap F_2$ is a $1$-good-neighbor conditional cut of $LTQ_n$, so $|F_1\cap F_2|\geq 2(n-1)$ by Lemma \ref{L6}. Since $\delta(LTQ_n[F_2-F_1])\geq 1$, we have $|F_2-F_1|\geq 2$. Therefore, $|F_2|=|F_2-F_1|+|F_1\cap F_2|\geq 2 +2(n-1)=2n$.

Next, we assume that $W\neq\emptyset$. Then, $|V(W_1)|=1$. If $F_1-F_2=\emptyset$, then $LTQ_n-F_1-F_2=LTQ_n-F_2$. The vertex in $W_1$ is one isolated vertex in $LTQ_n-F_2$, which contradicts the fact that $F_2$ is a $1$-good neighbor conditional faulty set. We suppose $F_1-F_2\neq\emptyset$. Arbitrarily choose a vertex $w\in W$. Then, $N(w)\subseteq F_1\cup F_2$. Since $F_1$ and $F_2$ are indistinguishable, $|N(w)\cap(F_2-F_1)|\leq 1$ and $|N(w)\cap(F_1-F_2)|\leq 1$ by Lemma \ref{L02}.
Owing to the fact that $F_1$ and $F_2$ are $1$-good-neighbor conditional faulty sets, we have $|N(w)\cap(F_2-F_1)|=|N(w)\cap(F_1-F_2)|=1$ and $|N(w)\cap(F_1\cap F_2)|=n-2\leq |F_1\cap F_2|$. Thus,
\begin{eqnarray*}
\sum _{w\in W}|N(w)\cap(F_1\cap F_2)|&=&|W|(n-2)\\
&\leq& \sum _{x\in F_1\cap F_2}d(x)\\
&\leq& |F_1\cap F_2|n\\
&\leq& (2n-2)n.
\end{eqnarray*}
It follows that $|W|\leq \frac{2n(n-1)}{n-2}=2n+\frac{2n}{n-2}\leq 2n+4$ when $n\geq 4$.

If $|V(W_k)|\leq 1$, then $V(LTQ_n)=W\cup(F_1\cup F_2)$. Thus, 
\begin{eqnarray*}
|F_1|+|F_2|&=& |V(LTQ_n)|+|F_1\cap F_2|-|W|\\
&\geq& 2^n+(n-2)-(2n+4)\\
&=& 2^n-n-6.
\end{eqnarray*}
Therefore, for $n\geq5$, we have
\begin{eqnarray*}
\max\{|F_1|, |F_2|\}&\geq& \left\lceil\frac{|F_1|+|F_2|}{2}\right\rceil\\
&\geq& 2^{n-1}-\left\lfloor\frac{n}{2}\right\rfloor-3\\
&\geq& 2n.
\end{eqnarray*}

Now, we assume that $|V(W_k)|\geq 2$. Note that $F_1$ and $F_2$ are $1$-good-neighbor conditional faulty sets. Thus, $F_1\cap F_2$ is a $1$-good-neighbor conditional cut of $LTQ_n$. By Lemma \ref{L6}, we have $|F_1\cap F_2|\geq 2(n-1)$. 
From the assumption that $|F_1\cap F_2|\leq 2n-2$, it follows that $|F_1\cap F_2|= 2n-2$. If $|F_1-F_2|\geq 2$ or $|F_2-F_1|\geq 2$, then we have $|F_1|\geq 2n$ or $|F_2|\geq 2n$. Now, assume $|F_1-F_2|=|F_2-F_1|=1$.
Suppose $F_1-F_2=\{v\}$ and $F_2-F_1=\{u\}$. Note that each vertex in $W$ is adjacent to both $u$ and $v$. By the definition of $LTQ_n$, there are at most two common neighbors for any pair of vertices in $LTQ_n$, from which it follows that $|W|\leq 2$.

\begin{figure}[hptb]
   \centering
  \includegraphics[width=8cm]{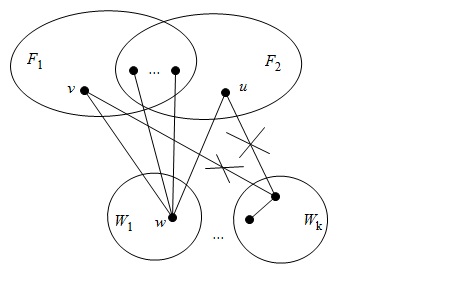}\\
  \caption{ The illustration of the proof of $|W|=1$.
}\label{p41}
\end{figure}

If $|W|=1$, suppose that $W=\{w\}$ and $N(w)-\{v, u\}\subseteq F_1\cap F_2$ (see Figure \ref{p41}). Since $LTQ_n$ contains no triangle by Lemma \ref{L7}, it follows that $\big(N(w)-\{v, u\}\big)\cap \big(N(v)-\{w\}\big)=\emptyset$ and $\big(N(w)-\{v, u\}\big)\cap \big(N(u)-\{w\}\big)=\emptyset$. Note that $|W|=1$ and $|F_1-F_2|=|F_2-F_1|=1$. By the fact that there is no edge between $W_i$ and $F_1\bigtriangleup F_2$ when $|V(W_i)|\geq 2$, we have $N(v)-\{w\}\subseteq F_1\cap F_2$ and $N(u)-\{w\}\subseteq F_1\cap F_2$. By the definition of $LTQ_n$, there are at most two common neighbors for any pair of vertices in $LTQ_n$, from which it follows that $|\big(N(v)-\{w\}\big) \cap\big(N(u)-\{w\}\big)| \leq1$. Thus,
\begin{eqnarray*}
 && |F_1\cap F_2|\\
&\geq& |N(w)-\{v, u\}|+|N(v)-\{w\}|+|N(u)-\{w\}|-1\\
&=& (n-2)+(n-1)+(n-1)-1\\
&=& 3n-5.
\end{eqnarray*}
It follows that $|F_2|=|F_2- F_1|+|F_1\cap F_2|\geq 1+3n-5=3n-4\geq 2n$ for $n\geq5$.

\begin{figure}[hptb]
\centering
  \includegraphics[width=8cm]{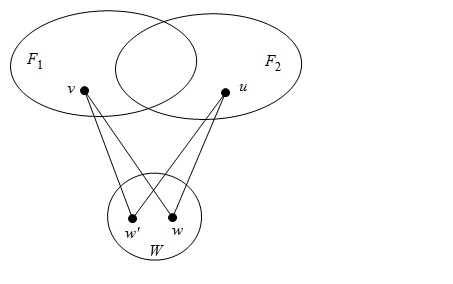}\\
  \caption{ The illustration of the proof of $|W|=2$.
}\label{p42}
\end{figure}

If $|W|=2$, suppose that $W=\{w, w'\}$ (see Figure \ref{p42}). Then both $v$ and $u$ are adjacent to $w$ and $w'$. Since $LTQ_n$ contains no triangle by Lemma \ref{L7} and there are at most two common neighbors for any pair of vertices in $LTQ_n$, we know that the four vertex sets $N(w)-\{v, u\}$, $N(w')-\{v, u\}$, $N(v)-\{w, w'\}$ and $N(u)-\{w, w'\}$ do not pairwise intersect. Therefore,
\begin{eqnarray*}
 && |F_1\cap F_2|\\
&\geq& |N(w)-\{v, u\}| +|N(w')-\{v, u\}| +|N(v)-\{w, w'\}|\\
&&+|N(u)-\{w, w'\}|\\
&=& (n-2) +(n-2) +(n-2) +(n-2)\\
&=& 4n-8.
\end{eqnarray*}
It follows that $|F_2|=|F_2-F_1|+|F_1\cap F_2|\geq 1+4n-8=4n-7\geq 2n$ for $n\geq5$.

The proof of this lemma is complete. $\qed$

Finally, we give the proofs of our main theorems.

\begin{thm}\label{main1}
Let $n$ be an integer with $n\geq 4$. Then, the $g$-good-neighbor conditional diagnosability of $LTQ_n$ under the PMC model is
\begin{equation*}\label{}
t_g(LTQ_n)=
\left\{
  \begin{array}{lll}
   2^g(n-g+1)-1, & \hbox{$1\leq g\leq n-3$}; \\
    2^{n-1}-1, & \hbox{$n-2\leq g\leq n-1$.}
  \end{array}
\right.
\end{equation*}
\end{thm}
\proof If $1\leq g\leq n-3$, by combining Lemma \ref{p1} and Lemma \ref{p3}, we have $t_g(LTQ_n)= 2^g(n-g+1)-1$.

Suppose that $n-2\leq g\leq n-1$. By Lemma \ref{L3}, $t_g(LTQ_n)\geq t_{n-3}(LTQ_n)$. Since $t_{n-3}(LTQ_n)=2^{n-1}-1$ and  $t_g(LTQ_n)\leq 2^{n-1}-1$ by Lemma \ref{p2}, we have $t_g(LTQ_n)= 2^{n-1}-1$.

This completes the proof of Theorem \ref{main1}. $\qed$

\begin{thm}\label{main2}
Let $n$ be an integer with $n\geq 5$. Then, the $g$-good-neighbor conditional diagnosability of $LTQ_n$ under the MM$^*$ model is
\begin{equation*}\label{}
t_g(LTQ_n)=
\left\{
  \begin{array}{lll}
   2^g(n-g+1)-1, & \hbox{$1\leq g\leq n-3$}; \\
    2^{n-1}-1, & \hbox{$n-2\leq g\leq n-1$.}
  \end{array}
\right.
\end{equation*}
\end{thm}
\proof If $1\leq g\leq n-3$ and $n\geq 5$, by combining Lemma \ref{p1} and Lemma \ref{p4}, we have $t_g(LTQ_n)= 2^g(n-g+1)-1$.

Suppose that $n-2\leq g\leq n-1$ and $n\geq 5$. By Lemma \ref{L3}, $t_g(LTQ_n)\geq t_{n-3}(LTQ_n)$. Since $t_{n-3}(LTQ_n)=2^{n-1}-1$ and  $t_g(LTQ_n)\leq 2^{n-1}-1$ by Lemma \ref{p2}, we have $t_g(LTQ_n)= 2^{n-1}-1$.

This completes the proof of Theorem \ref{main2}. $\qed$

\section{Conclusions}\label{6}
In this paper, we determine the $g$-good-neighbor conditional diagnosability of the $n$-dimensional locally twisted cube under the PMC model and the MM$^*$ model. We show that when $n\geq 4$, the $g$-good-neighbor conditional diagnosability of $LTQ_n$ under the PMC model is
\begin{equation*}\label{}
t_g(LTQ_n)=
\left\{
  \begin{array}{lll}
   2^g(n-g+1)-1, & \hbox{$1\leq g\leq n-3$}; \\
    2^{n-1}-1, & \hbox{$n-2\leq g\leq n-1$,}
  \end{array}
\right.
\end{equation*} and when $n\geq 5$, the $g$-good-neighbor conditional diagnosability of $LTQ_n$ under the MM$^*$ model is
\begin{equation*}\label{}
t_g(LTQ_n)=
\left\{
  \begin{array}{lll}
   2^g(n-g+1)-1, & \hbox{$1\leq g\leq n-3$}; \\
    2^{n-1}-1, & \hbox{$n-2\leq g\leq n-1$.}
  \end{array}
\right.
\end{equation*}
Future research on this topic will involve studying the $g$-good-neighbor conditional diagnosability of many network topologies. 

\section*{Acknowledgments}
The work is supported by the National Natural Science Foundation of China (11571044, 61373021) and the Fundamental Research Funds for the Central Universities.

\noindent

\end{document}